\documentclass{phc-proc4-auth}

\usepackage{epsfig}

\usepackage{amssymb}

\begin{document}

\begin{frontmatter}



\title{Pseudogap and Central Peak in the Emery Model\thanksref{proj}}
\thanks[proj]{Supported by the Croatian Government under Project 0119256.}

\author{D.K. Sunko and S. Bari\v si\'c}

\address{Department of Physics, Faculty of Science, University
        of Zagreb, Bijeni\v cka cesta 32, HR-10000 Zagreb, Croatia.}

\begin{abstract}
The effect of antiferromagnetic (AF) correlations is studied in the
framework of the three-band (Emery) model, with respect to experiments in
BSCCO. We study the pseudogap regime with a central peak. Detailed
dispersions of quasiparticle peaks show that one can simultaneously fit
Fermi surfaces and ARPES leading-edge energy scales. The band parameter
regime is a strong-coupling one: marked renormalization of the
copper-oxygen overlap, making it smaller than the oxygen-oxygen overlap,
while the copper-oxygen energy splitting is the largest of the three. The
same regime was found previously in a zeroth-order fit of Fermi surfaces.
The inclusion of AF correlations in a weak-coupling approach resolves the
only qualitative discrepancy of the zeroth-order mean-field slave-boson
calculation with experiment: it is argued that the observed large flat
region of the dispersion around the vH point is due to the very
non-dispersive central peak in the X-M direction. The sudden increase of
the experimental one-particle dispersion in the X-M direction is explained
by the quasiparticle strength shifting to the upper wing of the magnetic
pseudogap, as one moves further away from the X (van Hove) point. Near it,
the lower wing is predicted to be observed in the X-M direction, in
addition to the narrow central peak, giving rise to a two-peaked structure
below the Fermi level, as found experimentally.
\end{abstract}

\begin{keyword}
strongly correlated electrons \sep pseudogap \sep ARPES \sep high
temperature superconductors
\PACS 71.27+a \sep 74.72-h \sep 79.60-i
\end{keyword}
\end{frontmatter}

\section{Introduction}
\label{}

To understand conducting electrons in high-T$_c$ superconductors, one can
try and separate the experimental situation into as many individually
understandable pieces as possible. Simple calculations are then used to
read experiment, rather than predict it, and so constrain the eventual
complete theory. In this spirit, our main result here is that the narrow
non-dispersive feature found in ARPES measurements in BSCCO along the
$(\pi,0)$--$(\pi,\pi)$ (X--M) direction is antiadiabatic: the responsible
fermionic excitation is much slower than the dominant perturbing boson,
which imparts it with a $(\pi,\pi)$ momentum transfer.

This interpretation is obtained in the context of two other insights.
First, the observed Fermi surface shapes in LSCO, YBCO and BSCCO are most
efficiently fitted in a three band model, taking the oxygen degree of
freedom explicitly into account~\cite{Mrkonjic2001}. This reduces the
number of parameters needed from six (in a one-band model) to only three.
Second, while the extended high ARPES background is not obtained here, we
note it can be obtained by an explicit treatment of the on-site repulsion
by slave-bosons with fluctuations, both in ARPES~\cite{Melo1990} and Raman
spectra~\cite{Niksic2001}.

In this paper, we concentrate on the next piece of the puzzle. In BSCCO,
the above-mentioned Fermi surface fit is not accompanied by a similarly
successful fit of the dispersion. The experimental dispersion is very flat
in the X--M direction, while the zeroth-order dispersion of the Cu-O
resonant band, fitted to the Fermi surface, has a strong anisotropy at the
X (vH) point. We show that the discrepancy is resolved by the intervention
of a narrow antiadiabatic quasiparticle near the Fermi energy. The leading
edge scale of the wide adiabatic wing beneath it can be obtained at the
same time.

\section{Antiadiabatic regime}

We calculate the one-loop contribution to the electron self-energy,
extending a previous calculation~\cite{Sunko1993} to a realistic dispersion, and electrons
at arbitrary wave-vectors. The dispersion accounts for the strong on-site
repulsion through a mean-field renormalization of overlaps, so the
parameter regime is $t<|t'|<<\Delta_{pf}$, the \emph{effective} Cu-O
hopping, O-O hopping, and Cu-O energy splitting, respectively. The main
feature of this work is that the perturbing paramagnon is \emph{faster}
than the electron around the vH point. Such an antiadiabatic regime always
exists, because the electron velocity is zero there, but in our case, the
large quasiparticle effective mass and short correlation length of
paramagnons act in concert to extend the antiadiabatic quasiparticle to
about 30\% of the zone in the X--M direction. This requires a low, but not
too low `bandhead' of the paramagnons, we take 10 meV, appropriate for the
superconducting state~\cite{Mignod1991}.

\begin{figure}[t]
\center{\epsfig{file=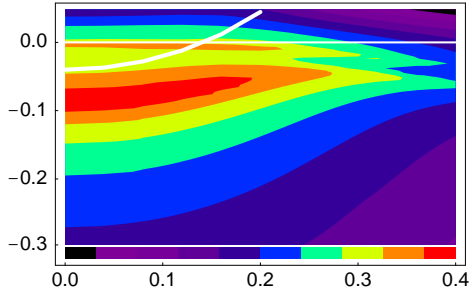,height=40mm}}
\caption{Log-intensity plot in the X--M direction, parameters of
the best fit to the Fermi surface: $t=0.4,t'= - 1.6,\Delta_{pf}=4.5,
\mu=-0.04$ eV. Note the effective flattening near $E_f$ due to the central
peak.The white line is the unperturbed band dispersion. Intensities are
multiplied by a Fermi factor. Axes are the same as in Fig.~\ref{exp}.}
\label{inten}
\end{figure}

The corresponding intensities are shown in Fig.~\ref{inten}. The net
effect on the dispersion around E$_f$ is as if one had cut out the rising
part and replaced it horizontally. The lower wing turns upwards, following
the unperturbed dispersion, but shifted by a pseudogap. The possible
impact of the side wings on high T$_c$ has recently been extensively
analysed~\cite{Friedel2002}. The antiadiabatic (horizontal) part has a
maximum where it crosses the trace of the old dispersion. The valley
separating the two peaks is due to the absence of explicit slave boson
dynamics in the calculation~\cite{Melo1990}.

\begin{figure}[t]
\center{\epsfig{file=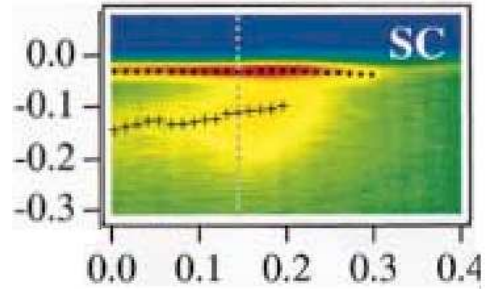,height=40mm}}
\caption{Experimental log-intensities in the X--M
direction~\cite{Kaminski2001}. The vertical scale is in eV from E$_f$,
the horizontal in $\pi/a$ away from the vH point.}
\label{exp}
\end{figure}

Figure~\ref{exp} shows the experimental intensities in the X--M
direction~\cite{Kaminski2001}. We observe that the lower wing turns
upward, and that the non-dispersive feature has a maximum where the lower
wing `points' at it. This unexpected correspondence of qualitative details
with our calculation increases our confidence in the basic interpretation
of the non-dispersive feature.

\section{Discussion}

The antiadiabatic quasiparticle is a very robust phenomenon, once the
paramagnon anomaly increases at a sufficient rate out of its minimum at
$(\pi,\pi)$, relative to the electron dispersion's increase away from the vH
point. This opens a `window' in the BZ, where the fermions are slower than
the bosons, so there is no adiabatic suppression of the quasiparticle
strength. Then the relative strength of the central peak,
and the energy scale at which the wings appear, may be adjusted through
the paramagnon bandhead and coupling strength, respectively, without much
fine-tuning.

As one moves away from the vH point in the $\Gamma$ direction, the
non-dispersive feature 'melts' with the lower wing into a single
dispersion, both in our calculation and in experiment. However, while this
is complete by $(0.9\pi/a,0)$ in the calculation, the non-dispersive
feature is observed up to $(0.6\pi/a,0)$~\cite{Kaminski2001}, as also
found in cuts along the $\Gamma$--M line~\cite{Valla2001}. We hope that
a self-consistent calculation, allowing for charge and spin channels on
equal footing, might resolve this issue.



\end{document}